\documentstyle[amscd,amssymb,12pt]{amsart}
\input xypic \xyoption{curve}

\def\hcorrection#1{\advance\hoffset by #1 }
\def\vcorrection#1{\advance\voffset by #1 }

\vcorrection{-.7in}
\hcorrection{-.5in}

\newcommand{\B}[1]{{\bold#1}} 
\newcommand{\C}[1]{{\cal#1}} 
\newcommand{\D}[1]{{\Bbb#1}} 

\theoremstyle{plain}
\newtheorem{th}{Theorem}[section]
\newtheorem{cor}{Corollary}[section]
\newtheorem{lem}{Lemma}[section]
\newtheorem{prop}{Proposition}[section]

\theoremstyle{definition}
\newtheorem{defin}{Definition}[section]

\theoremstyle{definition}
\newtheorem{example}{Example}[section]

\theoremstyle{remark}
\newtheorem{rem}{Remark}[section]
\newtheorem{notation}{Notation}[section]

\numberwithin{equation}{section}

\begin{document}

\pagestyle{plain}
\addtolength{\footskip}{.3in}

\title{Perturbative Quantum Field Theory\\
 and \\
Configuration Space Integrals\\
}
\author{Lucian M. Ionescu}
\address{Department of Mathematics, Illinois State University, Normal, IL 61790-4520}
\email{lmiones@@ilstu.edu}
\keywords{$L_\infty$-algebra, configuration spaces, renormalization, star-product}
\subjclass{Primary:18G55; Secondary:81Q30,81T18}

\begin{abstract}
$L_\infty-$morphisms are studied from the point of view of perturbative quantum field theory,
as generalizations of Feynman expansions.
The connection with the Hopf algebra approach to renormalization is exploited \cite{CKren1,K1,K2}.
Using the coalgebra structure (Forest Formula),
the weights of the corresponding expansions are proved to be cycles of the DG-coalgebra 
of Feynman graphs. 
This leads to graph cohomology via the cobar construction.

The properties of integrals over configuration spaces are investigated.
The aim is to develop a cohomological approach in order 
to construct the coefficients of formality morphisms
using an algebraic machinery,
as an alternative to the analytical approach 
using integrals over configuration spaces.
These weights are prototypical ``Feynman integrals''.
The connection with renormalization is suggested.

A second example of ``Feynman integrals'', defined as state-sum model,
is investigated.
The connection with a related TQFT is mentioned,
supplementing the Feynman path integral interpretation of Kontsevich formula. 

A categorical formulation for the Feynman path integral quantization is sketched.
\end{abstract}

\maketitle

\tableofcontents


\section{Introduction}\label{S:intro}
$L_\infty$-morphisms may be represented as perturbation series over
a ``category'' of Feynman graphs, 
as Kontsevich showed in \cite{Kon1}.
In this article we study such ``Feynman-Taylor'' expansions,
and their physical interpretation, supplementing \cite{Cat1}.

Applying the cobar construction to graph homology,
we prove that such $L_\infty$-morphisms correspond to the 
cohomology classes of the corresponding 
DG-coalgebra of Feynman graphs (Theorem \ref{T:ft}).

Example of such cocycles (periods/weights) are provided by integrals 
(``Feynman integrals'' corresponding to a given propagator and a class of Feynman graphs),
over the boundary of the compactification of configuration spaces of a given
manifold with boundary.

Part of the motivation is provided by our hope that 
algebraic examples of such cycles may be constructed based 
on the Hopf algebra of trees.
Applied to deformation quantization, 
this algebraic approach would provide the coefficients of the 
(local) star-product of a Poisson manifold.

The present approach is an alternative to the one using analytic tools to compute 
the coefficients of the deformation series,
namely using integrals over configuration spaces.
It reinforces the statement that renormalization is essentially an algebraic process,
independent on the regularization and renormalization schemes \cite{CKren1,K2}. 

This work is based on an analysis of \cite{Kon1},
aiming to extract an axiomatic ``interface'' from 
Kontsevich implementation of the formality morphism, 
which is based on integrals over configuration spaces.
In \cite{Kon1} the formula for the star-product is the mathematical implementation
of an open string theory (as declared in \cite{Kon1}, p.4, and announced in \cite{Kfc}, p.147), 
namely a Poisson sigma model on the disk \cite{Cat1}.
The coefficients of the terms of the 2-point functions are integrals
over compactified configuration spaces.
Reminiscent of regularization of Feynman path integrals, 
some intrinsic properties are extracted, 
and the algebraic condition establishing the $L_\infty$-morphism 
is reinterpreted as a certain ``Forest Formula'' (\cite{CKren1,K2}),
by using the coalgebra structure of renormalization.

As already stated, 
the main property of the compactification of configuration spaces 
is the coalgebra structure of their boundaries.
As a consequence, the periods of a closed form over the 
codimension one boundary define a cocycle of the cobar DG-algebra of Feynman
graphs.
Since these cocycles correspond to $L_\infty$-morphisms,
the success of using integrals over configuration spaces to implement
formality morphisms, and in particular deformation quantization formulas,
is explained.

It is known that the cobar construction 
allows us to compute the homology of loop spaces (\cite{A}, p81).
Its appearance in the process of understanding the ``loop structure'' of Feynman graphs
is hardly surprising.
An intrinsic approach to their study,
in the context of $A_\infty$-algebras,
is from the perspective of codifferentials on the tensor coalgebra,
suggesting the application of the cobar construction to graph homology.

We hope that our approach will contribute to the understanding of the
cycles obtained from an $A_\infty$-algebra \cite{P}, p.13,
also providing an instance of the plethora of ``partition functions'' obtained via a 
``state-sum model'' on a ``generalized cobordism category''
(\cite{INSF}; see also loc. cit. p.14).

%
%
The paper is organized as follows.
The results on $L_\infty$-morphisms expanded as a perturbation series over a 
class of ``Feynman'' graphs are explained in \S\ref{S:li}.
Their coefficients satisfy a certain cocycle condition, 
representing modulo homotopy the cohomology class of the
DG-coalgebra of Feynman graphs.

A mathematical interface to perturbative QFT is proposed supported by 
the findings of the next section.

A reader interested in the motivation for the above results may benefit 
from reading \S\ref{S:ics} first,
where the integrals over configuration spaces used in \cite{Kon1} are studied,
extracting some of their intrinsic properties,
notably the ``Forest Formula'' \ref{E:configb} of the boundary of their compactification.

Feynman state spaces and configuration functors are defined (graph cohomology).
Feynman rules defined as multiplicative Euler-Poincare maps are proved compatible with
the Forest Formula via the (Feynman) integration pairing (Theorem \ref{T:ff}).

The state-sum model from \cite{Kon1},
and yielding an example of a generalized Feynman integral,
is studied in Section \ref{S:ssc}.
The corresponding TQFT provides the additional evidence towards 
its interpretation as a Feynman path integral,
conform \cite{Cat1}.
The properties of the generalized Feynman rule
needed for the proof of the formality conjecture
are identified.

Section \ref{S:appl} applies the previous results to formality of DGLAs.
Kontsevich proof is translated into the language developed so far,
providing the essential steps for a proof of the main theorem from Section \ref{S:li}.

Section \ref{S:c} concludes with comments on the use of homotopical algebra 
towards the implementation of perturbative quantum field theory.

%
%
The main concern in this article, a project in its ``designing phase'',
is on mathematical concepts and their relations to QFT.
We thus apologize for being sketchy in several instances,
and deferring several details (and proofs) to another stage of development 
(``implementation stage'').

\vspace{.1in}
{\bf Acknowledgments}
I would like to express my gratitude for the excellent research conditions at I.H.E.S.,
where this project was conceived under the influence of,
and benefitting from stimulating discussions with Maxim Kontsevich.

Stasheff's comments are greatly appreciated. 
Critical comments from the reader (if any) will be equally appreciated.

\section{$L_\infty$-morphisms as perturbation series}\label{S:li}
We investigate when a graded map between $L_\infty$-algebras 
represented as a {\em Feynman expansion} over a given class of graphs 
(``partition function'') is an $L_\infty$-morphism.
The goal is to understand the coefficients of formality morphisms 
and Kontsevich deformation quantization formula, 
as well as perturbative QFT (see \S \ref{S:c} for details).
We will prove that the obstruction for a pre-$L_\infty$-morphism (\cite{Kfc}, p.142)
to be a morphism is of cohomological nature,
and point to its relation with renormalization.

%
%
\subsection{Feynman graphs}\label{SS:FG}
Consider a graded class of {\em Feynman graphs} 
(e.g. 1-dimensional CW-complexes or finite graded category, 
i.e. both objects and Homs are finite in each degree),
and $g$ the vector space over some field $k$ of characteristic zero,
with homogeneous generators $\Gamma\in \C{G}_n$ (e.g. ``admissible graphs'' \cite{Kon1}, p.22).

A Feynman graph will be thought off both as an object in a 
category of Feynman graphs (categorical point of view),
as well as a cobordism between their {\em boundary vertices} (TQFT point of view).
The main assumption needed is the existence of subgraphs and quotients.

While the concept of {\em subgraph} $\gamma$ of $\Gamma$ 
is clear (will be modeled after that of a subcategory),
we will define the {\em quotient} of $\Gamma$ by the subgraph $\gamma$ 
as the graph $\Gamma'$ obtained by collapsing $\gamma$ (vertices and internal edges) 
to a vertex of the quotient (e.g. see \cite{CKren1}, p11).
\begin{rem}\label{R:quotient}
When $\gamma$ contains ``external legs'', 
i.e. edges with 1-valent vertices belonging to 
the boundary of the Feynman graph when thought of as a cobordism,
we will say that $\gamma$ {\em meets the boundary} of $\Gamma$.
In this case the vertex of the quotient obtained by collapsing $\gamma$
will be part of the boundary (of $\Gamma/\gamma$) too.
In other words the boundary of the quotient is the quotient of the boundary
(compare \cite{Kon1}, p.27).

Formal definitions will be introduced elsewhere.
\end{rem}
\begin{defin}\label{D:normal}
A subgraph $\gamma$ of $\Gamma\in\C{G}$ is {\em normal} iff the corresponding
{\em quotient} $\Gamma/\gamma$ belongs to the same class of Feynman graphs $\C{G}$.
\end{defin}
\begin{defin}\label{D:extension}
An {\em extension} $\gamma\hookrightarrow \Gamma\twoheadrightarrow \gamma'$ in $\C{G}$
is a triple (as displayed) determined by a subgraph $\gamma$ of $\Gamma$,
such that the quotient $\gamma'$ is in $\C{G}$.
The extension is a {\em full extension} if $\gamma$ is a {\em full subgraph},
i.e. together with two vertices of $\Gamma$ contains all the corresponding connecting arrows
(the respective ``Hom'').
\end{defin}
Edges will play the role of simple objects. 
\begin{defin}\label{D:sg}
A subgraph consisting of a single edge is called a {\em simple subgraph}.
\end{defin}
%
\begin{example}\label{Ex:graphs}
As a first example consider the class $\C{G}_a$ of {\em admissible graphs} provided in \cite{Kon1}.
Denote by $\C{G}$ the larger class of graphs,
including those for which edges from boundary points may point towards internal vertices 
(essentially all finite graph one-sided ``cobordisms'': $\emptyset\to[m]$).
Then the normal subgraphs relative to the class $\C{G}_a$ are precisely the subgraphs
with no ``bad-edge'' (\cite{Kon1}, p.27), 
i.e. those for which the quotient is still an admissible graph.

Another example is the class of Feynman graphs of $\phi^3$-theory.
In this context a subgraph of a 3-valent graph collapses to a 3-valent vertex 
precisely when it is a normal subgraph in our sense.
\end{example}
There is a natural pre-Lie operation based on the operation of insertion 
of a graph at an internal vertex of another graph \cite{K2}, \cite{CKins}, p14, (addressed next).
It is essentially a sum over extensions of two given graphs.
\begin{defin}\label{D:extprod}
The {\em extension product} $\star:g\otimes g\to g$ is the bilinear operation
which on generators equals the sum over all possible extensions of one graph
by the other one:
\begin{equation}
\gamma'\star \gamma=\sum_{\gamma\to\Gamma\to \gamma'}\pm \Gamma.
\end{equation}
\end{defin}
It is essentially the ``superposition of $Hom(\gamma,\gamma')$''.
As noted in \cite{CKins}, p.14, it is a pre-Lie operation,
endowing $g$ with a canonical Lie bracket (loc. cit. $\C{L}_{FG}$).

Let $H=T(g)$ be the tensor algebra with (reduced) coproduct:
\begin{equation}\label{E:fgcoprod}
\Delta\Gamma=\sum_{\gamma\to \Gamma\to \gamma'}\gamma\otimes\gamma',
\end{equation}
where the sum is over all non-trivial subgraphs of $\Gamma$ (``normal proper subobjects'')
such that collapsing $\gamma$ to a vertex yields a graph from the given class $\C{G}$ 
(compare with condition (7) \cite{CKren1}, p.11).
%
\begin{rem}
The two operations introduced are in a sense ``opposite'' to one another, 
since the coproduct unfolds a given graph into its constituents,
while the product assembles two constituents in all possible ways.
For the moment we will not dwell on the resulting algebraic structure.
\end{rem}
%
With the appearance of a Lie bracket and a comultiplication, 
we should be looking for a differential (towards a DG-structure).

Consider the graph homology differential \cite{Kon2}, p.109:
\begin{equation}\label{E:fgdiff}
d\Gamma=\sum_{e\in E_\Gamma} \pm \Gamma/{\gamma_e},
\end{equation}
where the sum is over the edges of $\Gamma$, $\gamma_e$ is the one-edge graph,
and $\Gamma/{\gamma_e}$ is the quotient
(forget about the signs for now).
\begin{th}\label{T:Fdgca}
$(H,d,\Delta)$ is a differential graded coalgebra.
\end{th}
\begin{pf}
That it is a coalgebra results from \cite{CKren1}, p.12.
So all we need to prove is that $d$ is a coderivation:
$$\Delta d=(d\otimes id+id\otimes d)\Delta.$$
Comparing the two sides (with signs omitted):
\begin{align}
LHS&=\sum_{e\in\Gamma} \sum_{\bar{\gamma}\subset \Gamma/e\to \bar{\gamma}'}
\bar{\gamma}\otimes\bar{\gamma'}\\
&=\sum_{e\in\Gamma}(
\sum_{e/e\in\bar{\gamma}\subset\Gamma/e\to\bar{\gamma}'}\bar{\gamma}\otimes \bar{\gamma}'+
\sum_{e/e\notin\bar{\gamma}\subset\Gamma/e\to\bar{\gamma}'}\bar{\gamma}\otimes \bar{\gamma}'
),
\end{align}
and
\begin{align}
RHS&=\sum_{\gamma\subset\Gamma\to\gamma'}
(\sum_{e\in\gamma}\gamma/e\otimes \gamma'+\sum_{e\in\gamma'}\gamma\otimes \gamma'/e)\\
&=\sum_{e\in\Gamma}
(\sum_{e\in\gamma\subset\Gamma\to\gamma'}\gamma/e\otimes\gamma'+
\sum_{e\notin\gamma\subset\Gamma\to\gamma'}\gamma\otimes\gamma'/e),
\end{align}
with a correspondence uniquely defined by 
$e\in\gamma\to \bar{\gamma}$, i.e. $\bar{\gamma}=\gamma/e$ and 
$e\in\gamma'\to \bar{\gamma}'$, i.e. $\bar{\gamma'}=\gamma'/e$ respectively,
concludes the proof.
\end{pf}
The boundary of the codimension 1 strata of the configuration spaces (see \S\ref{S:ics})
suggests to consider its cobar construction $C(H)=T(s^{-1}\bar{H})$ 
(\cite{GLS}, p.366, \cite{M}, p.171).
Moreover, this is the natural set up for DG(L)A-infinity structures (e.g. \cite{Kel}).

The total differential is $D=d+\bar{\Delta}$, 
where the ``coalgebra part'' $\bar{\Delta}$ is the graded derivation:
\begin{equation}\label{E:cbcp}
\bar{\Delta}\Gamma=\sum_{\gamma\to \Gamma\to \gamma'}\gamma\otimes\gamma',
\end{equation}
corresponding to the reduced coproduct $\Delta$ given by equation \ref{E:fgcoprod}.
\begin{defin}\label{D:fcobar}
The cobar construction $(C(H),D)$ of the DG-coalgebra $(H,d,\Delta)$ of Feynman graphs is 
called the {\em Feynman cobar construction} on $\C{G}$.
\end{defin}
%
%

Taking the homology of its dual $(Hom_{Calg}(C(H),k),\delta)$ relative some field $k$,
with dual differential $\delta$,
yields $H^\bullet(H;k)$, 
the cohomology of the DG-coalgebra of Feynman diagrams $\C{G}$.
We will see in section \ref{S:linf} that it characterizes 
$L_\infty$-morphisms represented as Feynman expansions.

%
%
\subsection{Feynman-Taylor coefficients}\label{SS:FR}
Let $(g_1,Q_1)$ and $(g_2,Q_2)$ be $L_\infty$-algebras, 
and $f:g_1\to g_2$ a pre-$L_\infty$ morphism (\cite{Kon1}, p.11)
with associated morphism of graded cocommutative coalgebras 
$\C{F}_*:C(g_1[1])\to C(g_2[1])$,
thought of as the Feynman expansion of a partition function:
$$\C{F}_*=\sum \C{F}_n, \quad \C{F}_n(a)=\sum_{\Gamma\in \C{G}_n}<\Gamma,a>, \quad a\in g_1^n,$$
where the ``pairing'' $<\ ,\ >$ corresponds to a morphism $B:H\to Hom(g_1,g_2)$.
\begin{defin}\label{D:gfi}
A morphism $B:H\to Hom(g_1,g_2)$ is called {\em a generalized Feynman integral}.
Its value $<\Gamma,a>$ will be called a {\em Feynman-Taylor coefficient}.

Characters $W:H\to \B{R}$ act on Feynman integrals:
$$U=W\cdot B, \quad U(\Gamma)=W(\Gamma)B(\Gamma), \Gamma\in \C{G}.$$
\end{defin}
An example of a generalized Feynman integral is $U_\Gamma$ defined in \cite{Kon1}, p.23,
using the pairing between polyvector fields and functions on $\D{R}^n$.
An example of (pre)$L_\infty$-morphisms associated with graphs 
is provided by $U_n=\sum_{\Gamma\in \C{G}_n}W_\Gamma B_\Gamma$,
the formality morphism of \cite{Kon1}, p.24 (see \S\ref{S:ics} for more details).

Feynman integrals as pairings involving Feynman rules corresponding to propagators 
(the common value on all edges, e.g. $W_\Gamma$ in \cite{Kon1}, p.23)
will be defined in Section \ref{SS:fri} (Definition \ref{D:Frule}).

%
%
\subsection{$L_\infty$-morphisms}\label{S:linf}
Before addressing the general case of $L_\infty$-algebras, 
we will characterize formality morphisms of DGLAs 
(e.g. polyvector fields and polydifferential operators).
\begin{th}\label{T:f1}
Let $(g_1,0,[,]_{SN})$ and $(g_2,d_2,[,])$ be two DGLAs,
and $f=W\cdot U:g_1\to g_2$ a pre-$L_\infty$-morphism as above.
Then

(i) $\delta W=[f,Q]$,
where $Q$ denotes the appropriate $L_\infty$-structure.

(ii) $f$ is an $L_\infty$-morphism iff the character $W$ is a cocycle 
of the DG-coalgebra of Feynman graphs: $\delta W=0$.
\end{th}
The proof in the general case is essentially the proof from \cite{Kon1},
which will be given in section \ref{S:appl} (see Theorem \ref{T:dgla}).
\begin{defin}
A character $W$ is called a {\em weight} if it is such a cocycle.
\end{defin}
We claim that the above result holds for arbitrary $L_\infty-algebras$.
Moreover $L_\infty$-morphisms can be expanded over a suitable class of Feynman graphs,
and their moduli space corresponds to 
the cohomology group of the corresponding DG-coalgebra of Feynman graphs.
\begin{th}\label{T:ft}
(``Feynman-Taylor'')

Let $\C{G}$ be a class of Feynman graphs and $g_1, g_2$ two $L_\infty$-algebras.

In the homotopy category of $L_\infty$-algebras, 
$L_\infty$-morphisms correspond to the cohomology of the corresponding Feynman DG-coalgebra:
$$\C{H}o(g_1,g_2)=H^\bullet(H;k).$$
\end{th}
The basic examples (formality morphisms) are provided by cocycles 
constructed using integrals over compactification of configuration spaces 
(periods (\cite{Kon4}, p.26; see \S\ref{S:ics} and \S\ref{S:appl} for more details.)
%
\begin{rem}\label{R:framework}
The initial motivation for the present approach was 
to find an algebraic construction for such cocycles.
The idea consists in defining an algebraic version of 
the ``configuration functor'' $S:H\to C_\bullet(M)$,
a ``top'' closed form $\omega:H\to \Omega^\bullet(M)$ 
and a pairing $<S,\omega>$.
Their properties suggest the following framework, 
which will be detailed in section \ref{SS:fri}:
a chain map $S:(H,d)\to (C_\bullet,\partial)$,
and a cocycle $\omega$ in some dual cohomological complex $(C^\bullet(R)$:
$$<\partial S, \omega>=<S, d\omega>(=0),$$
so that the ``Stokes theorem'' holds. 
Then $W=<S,\omega>$ would be such a cocycle.
\end{rem}
A physical interpretation will be suggested here, 
and investigated elsewhere.

%
%
\subsection{A physical interpretation}\label{SS:pi}
Let $H$ be the Hopf algebra of a class of Feynman graphs $\C{G}$.
If $\Gamma$ is such a graph, 
then configurations are attached to its vertices,
while momenta are attached to edges 
in the two dual representations (Feynman rules in position and momentum spaces).

This duality is {\em represented} by a pairing between a ``configuration functor''
(typically $C_\Gamma$, see \S\ref{SS:bs}),
and a ``Lagrangian'' (e.g. $\omega$ determined by its value on an edge, i.e. by a propagator).
Together with the pairing (typically integration) representing the action,
they are thought of as part of the Feynman model of the 
{\em state space of a quantum system}.
\begin{rem}\label{R:pi}
As already argued in \cite{Irem},
this ``Feynman picture'' is more general than 
the manifold based ``Riemannian picture'',
since it models in a more direct way the observable aspects 
of quantum phenomena (``interactions'' modeled by a class of graphs),
without the assumption of a continuity (or even the existence) 
of the interaction or propagation process in an ambient ``space-time'', 
the later being clearly only an artificial model useful to relate with
the classical physics, i.e. convenient for ``quantization purposes''.
\end{rem}
\begin{defin}\label{D:action}
An {\em action} on $\C{G}$ (``$S_{int}$''),
is a character $W:H\to \B{R}$
which is a cocycle in the associated DG-coalgebra $(T(H^*),D)$.

A source of such actions is provided by a 
morphism of complexes $S:H\to C_\bullet(M)$ (``configuration functor''),
where $M$ is some ``space'', 
$C_\bullet(M)$ is a complex (``configurations on $M$''),
endowed with a pairing $\int:C_\bullet(M)\times\Omega^\bullet(M)\to \B{R}$,
where $\Omega^\bullet(M)$ is some dual complex (``forms on configuration spaces''),
i.e. such that ``Stokes theorem'' holds:
$$<\partial S,\omega>=<S, d\omega>.$$
A {\em Lagrangian} on the class $\C{G}$ of Feynman graphs 
is a k-linear map $\omega:H\to \Omega^\bullet(M)$ associating 
to any Feynman graph $\Gamma$
a closed volume form on $S(\Gamma)$ vanishing on the boundaries, 
i.e. for any subgraph $\gamma\to \Gamma$ (viewed as a subobject)
meeting the boundary of $\Gamma:[s]\to [t]$ (viewed as a cobordism),
$\omega(\gamma)=0$.

The associated action is $W=<S,\omega>.$
\end{defin}
A prototypical ``configuration functor''
is given by the compactification of configuration spaces $C_{n,m}$
described in \cite{Kon1} (see \S\ref{S:ics}).
The second condition for a Lagrangian emulates the vanishing on the boundary of 
the angle-form $\alpha$ (see \cite{Kon1}, p.22).
The coefficient $W(\Gamma)$ is then a {\em period} of the quadruple 
$(C_\Gamma,\partial C_\Gamma, \wedge_{k=1}^{|E_\Gamma|}\alpha(z_{i_k},z_{j_k})$
(\cite{Kon4}, p.24).
A related formulation (effective periods) is given in \cite{Kon4}, p.27.

\section{Integrals over configuration spaces}\label{S:ics}
The formality morphism $U$ from \cite{Kon1} was constructed
using ideas from string theory (loc. cit. p.1).
The terms of the ``n-point function'' $U_n$ are products of factors
determined by the interaction term (1-form on the disk) and the kinetic part
determining the propagator (Lagrangian ``decouples'').

With this interpretation in mind, we will investigate the properties of 
the $L_\infty$-morphism and its coefficients $W$ by analyzing 
the corresponding integrals on configuration spaces.
The coefficients $W(\Gamma)$ of the terms of the n-point function $U_n$ are expressed as
integrals over configuration spaces of points of a closed form vanishing on the boundary.

We claim that the main property of the compactification of the configuration space
(a manifold with corners),
is the ``Forest Formula'' (reminiscent of renormalization),
giving the decomposition of its boundary into disjoint strata.
This formula is a consequence of the fact that ``...open strata of $C_{n,m}$ are
naturally isomorphic to products of manifolds of type $C_{n',m'}$ and $C_{n'}$''
\cite{Kon1}, p.19.
The implications for the corresponding integrals and the properties of the integrands
stated in \cite{Kon1} are translated in our language 
targeting a categorical and cohomological interpretation.
Special consideration is given to the correspondence 
between the $L_\infty$-morphism condition $(F)$
and the coefficients $c_\Gamma$ of the Feynman expansion (see \cite{Kon1}, p.25).

%
%
\subsection{Configuration spaces}\label{S:cs}
Consider first the configuration space of n-points in a manifold $M$ (``no boundary'' case)
denoted by $C_n(M)$.
Then its compactification has the following structure:
$$\bar{C}_n(M)=\bigcup_{k=1}^{n-1}\ \bigcup_{k-forests}C_F,$$
where $C_F$ is a certain bundle over the configuration space of 
the roots of the forest $F$ with $k$ trees (\cite{Kon2}, p.106; \cite{Kon1}, p.20).
The codimension of a stratum equals $k$, 
the number of trees in the forest (\cite{BT}, p.5280).

We will be interested in the codimension one strata, 
for which Stokes theorem holds (see \cite{BT}, p.5281, (A3)).
This relevant part of the boundary of the configuration space,
denoted by $\partial \bar{C}_n(M)$, 
is a disjoint union of strata in one-to-one correspondence with
proper subsets $S\subset \{1,2,...,n\}=[n]$ with cardinality at least two (\cite{Kon2}, p.106):
\begin{equation}\label{E:strata}
\partial \bar{C}_n(M)=\bigcup_{[1]\subsetneqq S\subsetneqq[n]} \partial_S \bar{C}_n(M), \qquad 
\partial_S \bar{C}_n(M)\cong C_S\times C_{[n]/S}.
\end{equation}
On the right, the ``quotient'' of $[n]$ by the ``non-trivial subobject'' $S$
was prefered to the equivalent set with $n-|S|+1$ elements (``in the category of pointed sets'').

This formula involves a (reduced) coproduct structure,
the same way Zimmerman forest formula does,
in the (similar) context of regularized Feynman integrals and renormalization.

To extract the intrinsic properties of integrals over configuration spaces,
we will follow the proof of the formality theorem \cite{Kon1}, p.24, 
and record the relevant facts in our homological-physical interpretation:
admissible graphs are ``cobordisms'' $\Gamma:\emptyset\to [m]$ 
when $U_n$ is thought of as a state-sum model (\cite{Kon2}, p.100; see Section \ref{S:ssc}).
The graphs are also interpreted as ``extensions'' $\gamma\to \Gamma\to \gamma'$,
when considering the associated Hopf algebra structure.
The implementation of the concepts hinted above in quotation marks 
is scheduled for another article.

%
%
\subsection{Boundary strata of codimension one}\label{SS:bs}
Let $C_{n,m}$ be the configuration space of 
$n$ interior points and $m$ boundary points
in the manifold $M$ with boundary $\partial M$ 
(e.g. \cite{Kon1}, p.6: upper half-plane $\C{H}$).
Its elements will be thought of as (geometric) ``representations of cobordisms''
(enabling degrees of freedom with constraints):
$$\{\emptyset \overset{[n]}{\to} [m]\}\overset{x}{\longrightarrow} 
\{\emptyset\overset{M}{\to} \partial M\}.$$
$C_{n,m}$ will be also denoted by $C_{A,B}$, 
where $A$ and $B$ are the sets of internal respective boundary vertices,
with $n$, respectively $m$ elements.
This notation will extend to graphs $\Gamma$,
where $C_\Gamma=C_{A,B}$ will denote the space of states 
of the ``cobordism'' $\Gamma$ (see above Remark \ref{R:pi}),
at the level of vertices $\Gamma^{(0)}$ ($A/B$ the set of internal/boundary vertices).
\begin{defin}\label{D:fconfig}
The {\em space of configurations of $\Gamma$} is $C_\Gamma$,
the set of embeddings $\sigma$ of the set $\Gamma^{(0)}$ of its vertices
into the manifold with boundary $(M,\partial M)$,
which respects the boundary (``source'' and ``target''),
i.e. mapping {\em internal vertices} from $[n]$ to internal points of $M$,
and boundary vertices from $[m]$ to boundary points from $\partial M$.
\end{defin}
If $\gamma$ is a subgraph in $\Gamma$,
then a configuration of $\Gamma$ will induce by restriction 
a configuration on $\gamma$ (functoriality of configuration spaces; see \cite{BT}, p.5247).

Note that there is no canonical configuration 
induced on the corresponding quotient $\gamma'=\Gamma/\gamma$,
and this is where the compactification plays an important role.
Nevertheless the $M$-position of the vertex to which $\gamma$ 
collapses will belong to the boundary of $M$ iff 
the $\gamma$ meets the boundary $[m]$ of $\Gamma$ (see Remark \ref{R:quotient}).
The properties of these two distinct cases (``type S1/S2''),
will be treated below, and unified later on.

The codimension one boundary strata decomposes as follows (\cite{Kon1}, p.22):
\begin{equation}\label{E:configb}
\partial\bar{C}_{n,m}=(\bigcup_{S_1} \partial_{S_1} \bar{C}_{n,m}) 
\cup (\bigcup_{S_1,S_2}\partial_{S_1,S_2}\bar{C}_{n,m}).
\end{equation}
where $S_1$ and $S_2$ are subsets of points of $[n]$ and $[m]$.

Before briefly mentioning the tree-description of Equation \ref{E:strata},
we will reinterpret the above ``definition'' of the various portions of the
boundary $\partial_{S_1,S_2}\bar{C}_{n,m}$.
\begin{defin}\label{D:boundarymap}
Let $\Gamma$ be a graph with internal vertices $[n]$ and external vertices $[m]$
(cobordism $\Gamma:\emptyset\to [m]$, $[n]=\Gamma^{(0)}_{int}$).
Then $\partial_{S_1,S_2}\bar{C}_\Gamma$ denotes the portion of 
the codimension one boundary of the compactification of $C_\Gamma$ ($\subset\bar{C}_\Gamma$)
corresponding to the above decomposition.
If $S_2=\emptyset$, it will also be denoted by $\partial_{S_1}\bar{C}_\Gamma$.

For a full subgraph $\gamma$ (Definition \ref{D:extension})
determined by the sets of vertices $S_1$ and $S_2$,
the codimension one stratum $\partial_{S_1,S_2}\bar{C}_\Gamma$ 
will also be denoted as $\partial_\gamma\bar{C}_\Gamma$.
\end{defin}
Whether $\gamma$ intersects the boundary or not ($S_2=\emptyset$),
$S=S_1\cup S_2$ denotes the vertices which in the process of 
completion of the configuration spaces yield ``Cauchy sequences''
with $M$-coordinates getting closer and closer to one another
(see \cite{Kon1}, p.20 for additional details).

If $n_2=|S_2|\ge 2$,
the corresponding stratum can be alternatively labeled by the following tree:
$$
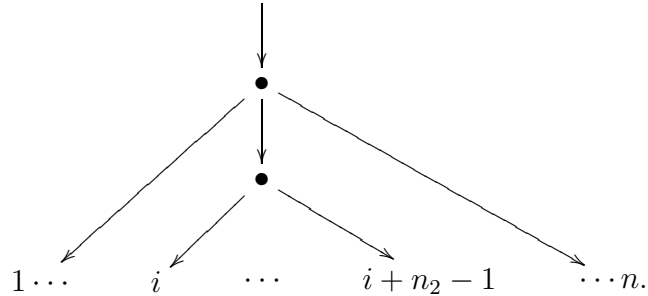
\diagram
&  & \dto^{} \\
&  & \bullet \ddllto^{} \dto^{} \ddrrto^{} & &\\
&  & \bullet \dlto^{} \drto^{} &  &\\
1 \cdots & i & \cdots & i+n_2-1 & \cdots n.
\enddiagram$$
It can be obtained by the insertion of the bottom line {\em $n_2$-corolla} \cite{Vor}, p.3,
in a leaf of the top line corolla, 
yielding a term of the graph homology differential \cite{Vor}, p.4.
Represented on an algebra, 
the operation becomes the $\circ_i$ Gerstenhaber composition (\cite{GS}; \cite{IHoch}, p.4),
yielding a term of the Hochschild differential.

Regarding codimension one strata, 
we will list the facts proved in \cite{Kon1}, p.25-27, 
using our notation aiming to generalize the usual context of Feynman graphs
to more general ``cobordism categories'' (\cite{INSF}).
Recall that the graphs considered are the ``admissible graphs'':
$\C{G}_a$ as defined in Example \ref{Ex:graphs}.
The results still hold for the class $\C{G}$,
which is closed under taking subquotients 
(A Serre category seems to be the appropriate context for considering 
Feynman integrands as Euler-Poincare maps - see Definition \ref{D:Frule}).
\begin{lem}\label{L:S1}
({\em ``Type\ $S_1$''})
If $\gamma_S$ is a non-trivial full subgraph in $\Gamma\in\C{G}_a$ 
supported on $S$ (set of vertices)
and not intersecting the boundary $[m]$ of $\Gamma$ (``vacuum fluctuation''),
then it is normal:
$$\gamma_S\hookrightarrow \Gamma\twoheadrightarrow \gamma',$$
with $\gamma'$ the corresponding quotient,
and the following ``short exact sequence is split'':
$$C_{\gamma}\hookrightarrow \partial_S\bar{C}_\Gamma\twoheadrightarrow C_{\gamma'},$$
i.e. $\partial_S\bar{C}_\Gamma=C_{\gamma_S}\times C_{\gamma'}.$
\end{lem}
\begin{pf}
Note first that a set of vertices $S$ determines uniquely a full subgraph $\gamma_S$.
Any such subgraph is ``normal'' as a ``subobject'',
i.e. the quotient is admissible, i.e. exists in the given class of Feynman graphs $\C{G}_a$.

The rest of the statement is a translation of the corresponding one in loc. cit. p.25.
\end{pf}
If the subgraph intersects the boundary, then the quotient may be a non-admissible graph 
(\cite{Kon1}, p.27: ``bad-edge'' sub case).
In all cases of ``type $S_2$'' we have the following:
\begin{lem}\label{L:S2} ({\em ``Type S2''})
If $\gamma_{S_1,S_2}$ is a non-trivial full subgraph in $\Gamma\in\C{G}_a$ 
supported on internal vertices from $S_1$ and 
intersecting the boundary $[m]$ of $\Gamma$ along $S_2$,
then $\gamma_{S_1,S_2}$ is normal in $\Gamma$ viewed as an object of $\C{G}$:
$$\gamma_{S_1,S_2}\hookrightarrow \Gamma\twoheadrightarrow \gamma',$$
with $\gamma'$ the corresponding quotient, 
and the following ``s.e.s is split'':
$$C_{\gamma_{S_1,S_2}}\hookrightarrow \partial_{S_1,S_2}\bar{C}_\Gamma\twoheadrightarrow C_{\gamma'},$$
i.e. $\partial_{S_1,S_2}\bar{C}_\Gamma=C_{\gamma_{S_1,S_2}}\times C_{\gamma'}$.
\end{lem}
\begin{pf}
Note first that an arbitrary set of vertices $S$ (here $S_1\cup S_2$), 
internal or not, determines uniquely a full subgraph $\gamma_S$.
Also recall that the boundary of the quotient of $\Gamma$ is also a quotient:
$\gamma':\emptyset\to [m]/S_2$ (see Remark \ref{R:quotient}).

loc. cit. p.26. 
\end{pf}
\begin{rem}
If $n_2=|S_1|$ and $m_2=|S_2|$, then the condition $n_2+m_2<n+m$ is equivalent to $S\subsetneqq \Gamma$,
and $2n_2+m_2-2\ge 2$ is equivalent to $|S|\ge 2$ under the assumption that $S=S_1\cup S_2$ intersects the
boundary $[m]$.
\end{rem}
\begin{rem}
The ``bad edge'' sub case (when the quotient is not an admissible graph),
is eliminated at the level of integration, yielding zero integrals.
Therefore from the point of view of the periods of a closed form,
one may restrict to consider only ``normal subobjects'', 
i.e. considering non-trivial ``extensions'' within the given class of Feynman graphs
realizing the given ``object'' $\Gamma$.

It will be proved elsewhere that this is the natural coproduct to be considered
in a suitable ``category of Feynman graphs''.
\end{rem}
%
In view of the above remarks and with the notation from Definition \ref{D:boundarymap},
the two lemmas may be summed up as follows.
\begin{prop}\label{P:Pext}
For any non-trivial full extension
$\gamma\hookrightarrow \Gamma\twoheadrightarrow \gamma'$ in $\C{G}$,
the following ``short exact sequence is split'':
$$C_\gamma\hookrightarrow \partial_\gamma\bar{C}_\Gamma\twoheadrightarrow C_{\gamma'},$$
i.e. $\partial_\gamma\bar{C}_\Gamma=C_{\gamma}\times C_{\gamma'}.$
\end{prop}
\begin{rem}
Note that from the point of view of integration,
edges with the same source and target will yield zero integrals,
since the ``propagator'' will be required to be zero on the diagonal:
$\Delta_F(x,x)=0$.
Alternatively, considering equivalence classes of graphs with orientation
(\cite{CV}, p.2) would eliminate the graphs with one edge loops.

One way or the other, all extensions may be considered.
\end{rem}

%
%
\subsection{Feynman state spaces}\label{SS:fri}
In this section we identify some intrinsic properties of 
integration of differential forms over the compactification of configuration spaces.
These properties will lead to cohomological statements relating them to the 
cobar DG-algebra of Feynman graphs.

Let $\Gamma\in\C{G}^l_{n,m}$ be a Feynman graph of {\em type $n,m$} and {\em degree $l$}, 
i.e. $\Gamma:\emptyset\to [m]$ with $n$ internal vertices, $m$ external legs
and $2n+m-2+l$ edges.
Then the compactification of the configuration space of $\Gamma$, $\bar{C}_\Gamma(M)$, 
is a manifold with corners, of dimension $k=2n+m-2$ (\cite{Kon1}, p.18).
Consider the rule $\Gamma\mapsto \omega(\Gamma)$, 
associating to a Feynman graph $\Gamma\in\C{G}^l_{n,m}$ 
the differential form on $C_{\Gamma}(M)$, of codimension $-l$
corresponding to the ``propagator'' $\Delta_F(x,y)=d\phi(x,y)$, 
where $d\phi$ is the angle form on $\C{H}$ as defined in \cite{Kon1}, p.6.
Note that the differential form $\omega(\Gamma)$
is closed and ``vanishing on the boundary'',
i.e. $d\phi(x,y)=0$ when $x\in\B{R}$.

To have a non-trivial integration pairing with $\partial C_\Gamma(M)$,
the codimension one strata of the boundary,
$\omega(\Gamma)$ must have codimension one too, 
i.e. $\Gamma$ should have $2n+m-3$ edges (\cite{Kon1}, p.25: $l=-1$).

The above $C$ and $\omega$, together with the integration pairing,
can naturally be extended to $H$ (see Section \ref{SS:FG}).
We will ignore for the moment the natural categorical setup,
where for instance $H$ is the Grothendieck ring of a strict monoidal category etc.

Note also that to account for orientations (ignored all together with signs for the moment), 
the vertices/edges of the Feynman graphs must be labeled.

\subsubsection{Feynman configurations}\label{SS:fc}
A ``simplicial (co)homology'' is considered,
with models the class of graphs $\C{G}$ (rather then trees, for now)
to play the role of the family of standard simplices $\{\Delta_n\}_{n\in \D{Z}}$.
Of course, the natural thing to do to obtain a genuine configuration functor,
would be to accept Feynman graphs for what they are: small categories.
We will postpone ``categorifying'' the Hopf algebra of Feynman graphs
(or rather discarding their decategorification),
and therefore, in order to map it to the corresponding configurations,
we will have to forget the internal structure of $C_\Gamma$,
by considering the Grothendieck ring of the category of configuration spaces.
%
%
\begin{defin}\label{D:configspace}
The category of {\em configurations of Feynman graphs in $M$},
denoted $\C{C}=\C{C}(\C{G},M)$,
is the additive strict monoidal category generated by the objects $C_\Gamma(M)$
with morphisms determined by equivalences of (labeled) Feynman graphs,
and tensor product $\times$ corresponding to disjoint union of graphs.
\end{defin}
The above propagator $\Delta_F=d\phi$ is determined by 
a function $\phi:C_e(M)\to \D{R}$ (element of $C_e(M)^*$: ``Lagrangian''),
where the simple object $e$ (the edge) is assumed to belong to $\C{G}$.

Consider the Grothendieck k-algebra of the above category,
``graded'' by $\C{G}$:
$$C_\bullet(M)=\D{R}\otimes K_0(\C{C}).$$

Since $H$ is the free (DG-co)algebra with generators $\Gamma$,
the embedding of generators map $C$ extends uniquely to $H$ as a k-algebra homomorphism 
$S:H\to C_\bullet(M)$ (an isomorphism!):
$$S(\sum_i\Gamma_i)=\oplus_i c(\Gamma_i),$$
$$S(\Gamma_1\cdot\Gamma_2)=c(\Gamma_1)\times c(\Gamma_2), \quad \Gamma_i\in\C{G},$$
where $(M)$ is tacitly understood,
$c_\Gamma$ denotes the isomorphism class of $C_\Gamma(M)$,
and the alternative notation for addition and multiplication in the target space is meant
to remind us about the categorical interpretation,

Now the category $\C{C}$ has a sort of a cone functor represented 
by the compactification functor:
$$C_\Gamma(M)\hookrightarrow\bar{C}_\Gamma(M).$$
%
The codimension one boundary of the compactification of the configuration spaces
has the following description:
\begin{equation}\label{E:boundary}
\partial \bar{C}(\Gamma)=
\sum_{\gamma\hookrightarrow \Gamma\twoheadrightarrow \gamma'}
\pm C(\gamma)\times C(\gamma'),
\end{equation}
where the sum is restricted to proper extensions.
It induces a derivation on the Grothendieck algebra
defined on generators as follows:
$$\partial c_\Gamma=[\partial \bar{C}_\Gamma(M)],$$
called the {\em boundary map}.
%
As wished for in Remark \ref{R:framework},
we have the following.
\begin{prop}
$(C_\bullet(M),\times,\partial)$ is a DG-algebra.
\end{prop}
The following essential property is a consequence of the definitions 
and of Proposition \ref{P:Pext}.
\begin{prop}\label{P:chainmap}
The $k$-algebra morphism $S:T(H)\to C_\bullet$ 
extending the configuration space functor $C$ is a chain map:
\begin{equation}\label{E:chainmap}
\partial S(\Gamma)=S(\Delta\Gamma),\quad \Gamma\in H.
\end{equation}
\end{prop}
\begin{pf}
It follows from definitions:
\begin{align}
\partial S(\Gamma)&=[\partial\bar{C}_\Gamma]=[\oplus_{\gamma\to\Gamma\to\gamma'}\ C_\gamma\times C_{\gamma'}]\\
&=\sum_{\gamma\to\Gamma\to\gamma'} S(\gamma)\times S(\gamma')=S(\Delta\Gamma).
\end{align}
Note that the right hand side involves 
the reduced coproduct defined by Equation \ref{E:cbcp}.
\end{pf}
The above proposition is taken as a defining property.
\begin{defin}\label{D:configfunctor}
A {\em configuration functor} is a DG-algebra morphism 
$S:(T(H),\otimes,\Delta)\to (C_\bullet, \times, \partial)$
from the Feynman cobar construction to the DG-algebra of {\em configuration spaces}.

Since equation \ref{E:boundary} will hold, a term $S(\gamma)\times S(\gamma')$,
corresponding to the subgraph $\gamma$ of $\Gamma$,
will be denoted by $\partial_\gamma S(\Gamma)$ (``face boundary maps'').
\end{defin}
\begin{rem}
The comparison with the simplicial case deserves some attention:
$$\partial_\Gamma=\sum_{\gamma\to\Gamma\to\gamma'} \pm \partial_\gamma, 
\quad \partial_n=\sum_{i=1}^n \partial_i,$$
including the analogy with the Eilenberg-Zilber maps \cite{DK}, p.55.
\end{rem}

%
%
\subsection{Feynman rules}
\begin{defin}\label{D:Frule}
A {\em Feynman rule} is a multiplicative Euler-Poincare map (see \cite{L}, p.98),
i.e. if $\gamma\hookrightarrow \Gamma\twoheadrightarrow \gamma'$ is an extension, then:
\begin{equation}\label{E:ext}
\omega(\Gamma)=\omega(\gamma)\wedge\omega(\gamma').
\end{equation}
The (common) value on an edge (a simple object) is called a {\em Feynman propagator},
and denoted by $\Delta_F=\omega(e)$.

A {\em Feynman integral} for the class of Feynman graphs $\C{G}$,
with Feynman rule $\omega(\Gamma)$ is
the k-algebra morphism extending $\omega$ to $H$, 
with values in $C^\bullet(M)=\Omega(C_\bullet(M))$
(Lagrangian \S\ref{SS:pi}; compare \cite{AS}, p.9):
$$\omega(\Gamma_1\cdot\Gamma_2)=\omega(\Gamma_1)\wedge\omega(\Gamma_2).$$
\end{defin}
Of course a Feynman integral is also a generalized Feynman integral according 
to Definition \ref{D:gfi}.

To justify the last part of the above definition,
recall that an Euler-Poincare map is determined by its values on simple objects,
and therefore a Feynman integrand has a common value on every edge (isomorphic objects).
Moreover, it descends on the corresponding Grothendieck algebra
(a normalization is assumed: $\omega(pt)=1$).
\begin{prop}\label{P:fr}
To any given propagator $\Delta_F$ there is a unique extension to a Feynman rule.
\end{prop}
\begin{pf}
Apply equation \ref{E:ext} for the case of simple subgraphs (edges).
\end{pf}
%
\subsection{Feynman integrals}
\begin{defin}\label{D:Fintegral}
The integration of forms $\omega(\Gamma)$ 
over the corresponding configuration space
$S(\Gamma)$ extends bilinearly, 
yielding a functional on $H$:
$$W(X)=\int_{S(X)}\omega(X), \quad X\in H.$$
This will be called a {\em Feynman integral} 
(action on $\C{G}$ - see \S\ref{SS:pi}, Definition \ref{D:action},
conform Theorem \ref{T:character} below).
\end{defin}
%
%
%
%
As expected $W$ is a character.
\begin{th}\label{T:character}
By extending $C$ and $\omega$ as algebra homomorphisms,
the natural pairing induced by integration:
$$W(X)=\int_{S(X)}\omega(X), \quad X\in H,$$
yields a character of the Hopf algebra of Feynman graphs.
\end{th}
\begin{pf}
This is essentially ``Fubini theorem''.
Indeed, on generators $\Gamma_i\in\C{G}, i=1,2$:
$$W(\Gamma_1\cdot\Gamma_2)=\int_{S_{(\Gamma_1\cdot\Gamma_2)}}\omega(\Gamma_1\cdot\Gamma_2)\\
=\int_{C_{\Gamma_1}\times C_{\Gamma_2}}\omega(\Gamma_1)\wedge\omega(\Gamma_2)\\
=W(\Gamma_1)\cdot W(\Gamma_2).
$$
\end{pf}
More important is the relation with the boundary map of the configuration functor
(considered next), as it will be shown later on.
It leads to the cohomological properties of the Feynman integrals.

%
%
\subsection{Cohomological properties of Feynman integrals}\label{SS:cp}
As a consequence of the previous Theorem \ref{T:character},
the integrals over the codimension one boundary match the codiferential of the 
cobar construction:
$$``\int_{\partial\bar{C}}=W\circ D''.$$
It follows that the character $W$ associated to the
configuration functor $S$ and propagator $\Delta_F$ is a cocycle
of the cobar construction of the Hopf algebra of Feynman diagrams.
This result will be used to prove the claim from Section \ref{S:li},
characterizing $L_\infty$-morphisms:
modulo equivalence they correspond to cohomology classes of the DG-coalgebra $H$.

Although the statements hold when the triple $(C,\bar{C},\partial)$ 
is replaced by any configuration functor $S$ (as the proofs show), 
to fix the ideas, assume $S$ extends the configuration functor $C$ and 
that $\omega=d\phi$ is the angle form as in \cite{Kon1}. 
Fix a Feynman graph $\Gamma\in\C{G}^{-1}_{n,m}$,
so that $\omega(\Gamma)$ is a codimension one form on the corresponding
configuration space $\bar{C}_\Gamma(M)$.

>From the previous results we deduce
the following translation of the statements from (6.4.1, 6.4.2) \cite{Kon1}, p.25,
regarding the integrals over the codimension one strata. 
\begin{prop}\label{P:multiplic}
Let $(C,\bar{C},\partial)$ be a configuration functor,
$\omega$ a Feynman rule with values in the algebra of differential forms
$\Omega(C_\bullet(M))$, 
and $W$ the Feynman integral corresponding to the natural pairing defined by integration.

For any full extension $\gamma\hookrightarrow \Gamma\twoheadrightarrow \gamma'$:
\begin{equation}\label{E:multiplic}
\int_{\partial_\gamma S(\Gamma)}\omega(\Gamma)=W(\gamma)\cdot W(\gamma').
\end{equation}
\end{prop}
\begin{pf}
By Proposition \ref{P:Pext} $\partial_\gamma\bar{C}_\Gamma=C_{\gamma}\times C_{\gamma'}.$
Since $\omega$ is an Euler-Poincare map (Definition \ref{D:Frule}),
the claim follows by ``Fubini theorem'':
$$\int_{\partial_\gamma\bar{C}_\Gamma}\omega(\Gamma)=
\int_{C_\gamma\times C_{\gamma'}}\omega(\Gamma)=
\int_{C_\gamma\times C_{\gamma'}}\omega(\gamma)\wedge\omega(\gamma')=
W(\gamma)\cdot W(\gamma').
$$
\end{pf}
\begin{rem}
Note that the statement holds also when $\gamma'$ is not admissible (``bad-edge'' case), 
due to the fact that the ``propagator'' $\Delta_F(x,y)$ vanishes on the boundary.
\end{rem}
Regarding the relation with the condition corresponding to $L_\infty$-algebra morphisms
((F) from \cite{Kon1}, p.24; see Section \ref{S:appl}),
note that some of the integrals over the codimension one boundary strata vanish, 
the remaining ones matching the coefficient $c_\gamma$ of $U_\Gamma$ 
in the Feynman expansion of (F).
$$``\int_{\partial\bar{C}}\omega=c''.$$
In order to distinguish various portions of the boundary $\partial\bar{C}$,
corresponding via Equation \ref{E:chainmap} to portions of the
reduced comultiplication of the Hopf algebra $H$,
we will introduce the following.
\begin{notation}
\begin{equation}\label{E:ecomult}
\Delta_e=\sum_{e\hookrightarrow\Gamma\twoheadrightarrow\Gamma/e} \pm \Gamma/e\otimes e.
\end{equation}
When restricted to internal edges, the above sum will be denoted by $\Delta_e^{int}$.
The corresponding portion of the sum in the graph homology differential \ref{E:fgdiff}
will be denoted by $d^{int}$.

In general, when considering internal subgraphs,
i.e. with their boundary consisting of internal vertices,
or subgraphs meeting the boundary,
the following notation will be used:
\begin{equation}\label{E:icomult}
\Delta_i=\sum_{\Gamma_2\hookrightarrow\Gamma\twoheadrightarrow\Gamma_1, 
\ \Gamma_2\cap\partial\Gamma=\emptyset} \pm \Gamma_1\otimes\Gamma_2.
\end{equation}
\begin{equation}\label{E:bcomult}
\Delta_b=\sum_{\Gamma_2\hookrightarrow\Gamma\twoheadrightarrow\Gamma_1, 
\ \Gamma_2\cap\partial\Gamma\ne\emptyset} \pm \Gamma_1\otimes\Gamma_2.
\end{equation}
$\Delta_{i-e}$ and $\Delta_{b-e}$
refer to sums over extensions where $\Gamma_2$ is not an edge
(all extensions are assumed to be proper).
\end{notation}
\subsubsection{Type S1 terms.}
First recall that any full subgraph not meeting the boundary $\partial M$
is a normal subgraph.
The integrals over the codimension one strata corresponding to such a 
subgraph $\gamma$ of type $(n,0)$,
i.e. not intersecting the boundary of $M$ (and yielding type $S_1$ terms),
with $n\ge 3$ vanish (see 6.4.1.2. \cite{Kon1}, p25).
 
The other terms ($n\le 2$) correspond to full subgraphs $\gamma$ 
consisting of one {\em internal edge} of $\Gamma$,
i.e. simple subgraphs not meeting the boundary.
The corresponding terms total the ``internal part'' of the 
graph homology differential \ref{E:fgdiff}.
\begin{prop}\label{P:s1}
For any Feynman graph $\Gamma\in\C{G}$:
$$\sum_{e\in\Gamma^{(1)}_{int}}\int_{\partial_e\bar{C}_\Gamma}\omega(\Gamma)=
W(d^{int}\Gamma),$$
where $e$ is a simple subgraphs of $\Gamma$, without boundary.
\end{prop}
\begin{pf}
Let $\gamma_e$ denote such a subgraph corresponding to the internal edge $e\in\Gamma^{(1)}$.
Applying Proposition \ref{P:multiplic} yields:
$$\sum_{e\in \Gamma^{(1)}}\int_{\partial_{\gamma_e}\bar{C}_\Gamma}\omega(\Gamma)\\
=\sum_{e\in \Gamma^{(1)}}W(\gamma_e)\cdot W(\Gamma/\gamma_e)\\
= coef.\cdot W(\sum_{e\in \Gamma^{(1)}}\pm \Gamma/\gamma_e)\\
= coef. \cdot W(d\Gamma),$$
where the coefficient depends only on the propagator $\Delta_F$.
With the appropriate normalization, we have:
$$W(\gamma_e)=\int_{\bar{C}_{2,0}(M)}d\phi_e=\pm 1$$
The sign is given by the labels of the two vertices of the internal edge,
corresponding to the convention for the orientation of graph homology.
\end{pf}
%
To emphasize the relation with the comultiplication from Section \ref{S:li},
the above results may be restated as follows.
\begin{cor}
$$W\circ\Delta_e^{int}=W\circ d^{int}, \quad W\circ\Delta_{i-e}=0.$$
\end{cor}

%
\subsubsection{Type S2 terms.}
Subgraphs $\Gamma_2$ which do meet the boundary of $\Gamma$ may produce 
quotients which are not admissible graphs ($\Gamma_1\notin\C{G}_a$).
\begin{lem}\label{L:badedge}
Feynman integrals over codimension one strata 
corresponding to non-normal subgraphs vanish.
\end{lem}
\begin{pf}
See \cite{Kon1}, p.27, the ``bad edge'' case 6.4.2.2.
\end{pf}
The remaining terms corresponding to normal proper subgraphs 
meeting the boundary $[m]$ of $\Gamma\in\C{G}_a$
yield a ``forest formula'' corresponding to the coproduct $\Delta_b$ of $\C{G}$.
\begin{prop}\label{P:s2}
For a Feynman graph $\Gamma\in\C{G}_a$:
\begin{equation}
\sum_{\gamma\hookrightarrow \Gamma\twoheadrightarrow\gamma'\ in\ \C{G}}
\int_{\partial_{\gamma}\bar{C}_\Gamma}\omega(\Gamma)=
W(\Delta_b\Gamma),
\end{equation}
where the proper normal subgraph $\gamma$ meets non-trivially the boundary of $\Gamma$.
\end{prop}
\begin{pf}
The formula follows from definitions and from the multiplicativity property of $W$ 
(Proposition \ref{P:multiplic}),
in the same way as for Proposition \ref{P:s1}.
\end{pf}
Putting together the two types of terms, $S_1$ and $S_2$,
and independent of the vanishing of some of the terms,
we obtain the following ``Forest Formula''.
\begin{th}\label{T:ff}
For any graph $\Gamma\in\C{G}$:
\begin{equation}\label{E:ff}
\int_{\partial\bar{C}_\Gamma}\omega(\Gamma)=W(\Delta\Gamma).
\end{equation}
\end{th}
\begin{rem}
The above result holds for an arbitrary configuration functor $S$ and
Feynman integrand $\omega$.
\end{rem}
So far we did not need $\omega(\Gamma)$ to be a closed form.
With this additional assumption,
using Stokes Theorem (duality at the level of a general configuration functor - 
see Remark \ref{R:framework}),
the closed form produces a cocycle.
\begin{cor}\label{C:Fcocycle}
If the ``Lagrangian'' $\omega$ is a closed form then the corresponding
Feynman integral $W$ is a cocycle.
\end{cor}
\begin{pf}
$$(\delta W)(\Gamma)=W(\Delta\Gamma)
\overset{T\ref{T:ff}}{=}\int_{\partial\bar{C}_\Gamma}\omega(\Gamma)
\overset{Stokes}{=}\int_{\bar{C}_\Gamma}d\omega(\Gamma)=0.$$
\end{pf}

The main property of the Feynman integrals $W$, the Forest Formula,
may be interpreted in a manner relevant to renormalization,
as follows.

Note first that $W$ is obtained as a ``cup product/convolution'':
\begin{equation}\label{E:convol}
S*\omega=\int\circ\ (S\otimes\omega)\circ c:H\to \B{R},
\end{equation}
where $c:H\to H\otimes H$ denotes the natural group-like comultiplication of $H$.
\begin{defin}\label{D:convolution}
The dual DG-algebra of {\em Feynman characters}:
$$H^*=\{w:H\to \B{R}|\ w \ character\ \}$$
is called the {\em convolution algebra} of $\C{G}$.
The corresponding differential is given by:
$$\delta w=w\circ \Delta, \quad w\in T(H)^*.$$
\end{defin}
\begin{th}\label{T:convolution}
$\delta$ is a derivation with respect to the ``convolution'' of 
the configuration functor $S$ and the Lagrangian form $\omega$:
$$\delta(S*\omega)=(\partial S)*\omega+S*(d\omega).$$
\end{th}
\begin{pf}
Let $W=S*\omega$ (Equation \ref{E:convol}).
Interpreting the left hand side of Equation \ref{E:ff} according to the above definition,
and using Theorem \ref{T:ff} yields:
$$(\delta W)(\Gamma)=W(\Delta\Gamma)=\int_{\partial S(\Gamma)}\omega(\Gamma).$$
Since a Lagrangian is a closed form, $d\omega=0$, the second term is zero,
and the equality is proved.
\end{pf}
The implications to the deformation point of view to renormalization 
will be considered elsewhere.

\section{State sum models and Feynman rules}\label{S:ssc}
In the previous section a Feynman rule with propagator $\Delta_F=d\phi$
paired via integration with a configuration functor $(C,\bar{C},\partial)$
produced the Feynman integral $W:H\to \D{R}$.

In this section a typical generalized Feynman rule is considered,
yielding the generalized Feynman integral 
$U:H\to Hom(T,D)$ of \cite{Kon1} (see Definition \ref{D:gfi}).
It is a typical ``state-sum model'' 
(state model \cite{Kon2}, p.100; see also \cite{T}, p.345),
having a Feynman path integral interpretation as already noted in \cite{Cat1}.

%
%
\subsection{A state-sum model on graphs}
We will review the construction mostly keeping the original notation.

For each graph $\Gamma\in\C{G}_a$, 
a function $\Phi=<U_\Gamma(\gamma),f>$ will be defined, 
where $\gamma=\gamma_1\otimes...\gamma_n$,
$f=f_1\otimes...\otimes f_m$ and $<,>$ denotes the natural evaluation pairing.

States of a graph have two conceptually distinct groups of data:
associating polyvector fields to internal vertices and appropriate functions to
boundary points.

First chose a basis $\{\partial_i\}_{i=1..d}$ for the algebra of vector fields.
A {\em coloring} of the labeled graph $\Gamma\in\C{G}_{n,m}$ (vertices are ordered),
is a map:
$$I:\Gamma^{(1)}\to \{1,...,d\}.$$
A {\em basic state} of the graph $\Gamma$ is the following association
corresponding to a coloring of its edges:
$$\phi^0(v)=\gamma_v, \ v\in\Gamma^{(0)}, \qquad
\phi^1(e)=\partial_{I(e)}, \ e\in\Gamma^{(1)}.$$
\begin{rem}
It is customary to implement $\phi^0$ via an ordering of the vertices of $\Gamma$,
obtaining the map $U_\Gamma:T^n\to Hom(A^m,A)$.

We prefered this more cumbersome notation (e.g. \cite{KS}, p.28)
because it reveals the true nature of a state-sum:
a 2-functor when interpreted categorically (\cite{INSF,Icatss}).
\end{rem}
Then $\Phi$ is the sum over all basic states 
corresponding to a fixed choice of $\phi^0$ ($\gamma's$ and $f's$),
of the corresponding ``amplitude'' (to be defined shortly):
$$\Phi=\sum_{all\ basic\ states\ \phi\ of\ \Gamma}\Phi_\phi.$$
\begin{rem}
The true amplitude of the process would involve a sum over all 
states, when the values of $\phi^0$ varies on the internal vertices
while the state of the boundary $f$ is fixed.
The sum over values on the edges amounts to a contraction process (traces etc.).
\end{rem}
Now $\Phi_\phi$ ($\Phi_I$ of \cite{Kon1}, p.23) is a product over the contributions
$\Phi_\phi(v)$ over the vertices of $\Gamma$, 
$n$ internal and $m$ boundary type.

For an internal vertex $v$:
$$\Phi_\phi(v)=(\prod_{e\in In(v)}\phi(e))\psi_v,\quad 
\psi_v=<\gamma_v,\bigotimes_{e\in Out(v)}dx^{I(e)}>,$$
where $In(v)$ ($Out(v)$) denotes the set of incoming (outgoing) edges of the vertex $v$,
and the shorthand notation $\phi=\phi^{(1)}$ was used 
since $\phi^0$ is fixed within this state-sum.

\subsection{The amplitude interpretation}
Towards an ``propagation amplitude'' interpretation, replace the evaluation pairing
with the inner product $(\ ,\ )$ such that the above basis $\{\partial_i\}_{i=1..n}$ 
be orthonormal.
Also collect the ``in'' and ``out'' products,
introducing the following terminology.
\begin{defin}\label{D:IOstates}
For any basic state $\phi$ of the graph $\Gamma$:
$$\phi_{In}=\prod_{e\in In(v)}\phi(e), \quad \phi_{Out}=\bigotimes_{e\in Out(v)}\phi(e),$$
are called the {\em In} and {\em Out} states of the scattering process at the vertex $v$.
\end{defin}
%
\begin{prop}\label{P:expval}
$$-\Phi_\phi(v)=(\phi_{Out}(v),ad_{\phi(v)}(\psi_{In}(v)))$$
is the {\em scattering amplitude}:
$$(\phi_{Out}|ad_{\phi}|\phi_{In})_{|v}$$
of the elementary process at the internal vertex $v$:
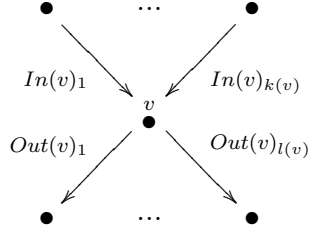
$$\diagram
\bullet \drto_{In(v)_{1}} & ... & \dlto^{In(v)_{k(v)}} \bullet \\
& \dlto_{Out(v)_1} \overset{v}{\bullet} \drto^{Out(v)_{l(v)}} & \\
\bullet  & ... & \bullet \\
\enddiagram$$
\end{prop}
\begin{pf}
Here $ad_X(Y)=[X,Y]$ is the commutation bracket on differential operators.
Therefore if $X$ and $Y$ commute, then $[X,fY]=X(f)Y$. 

Now to retrieve the appropriate component, use the above inner product:
$$\Phi_\phi(v)=\phi_{In}(\psi_v)=([\psi_{In}(v),\gamma_v],\phi_{Out}(v)).$$
This can be put in the form of a propagation amplitude $(\ |\ |\ )$,
establishing the above claim.
\end{pf}
%
In a similar manner, for boundary vertices we have the following
\begin{prop}
If $v\in\partial\Gamma$ is a boundary vertex, then:
$$\Phi_\phi(v)=<\phi_{In}(v),\phi(v)>,$$ 
is the ``expectation value'' of the process:
$$
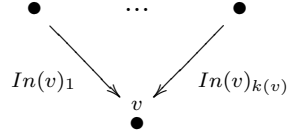
\diagram
\bullet \drto_{In(v)_{1}} & ... & \dlto^{In(v)_{k(v)}} \bullet \\
& \overset{v}{\bullet} & 
\enddiagram$$
where $<\ ,\ >$ denotes the natural evaluation pairing
between polyvector fields and functions.
\end{prop}
%
%
\subsection{A TQFT interpretation}\label{S:TQFT}
The properties of the generalized Feynman (path) integral $U$
may be viewed as consequences from the generalized TQFT 
implemented via the above state-sum model.
We will only sketch some points related to this TQFT interpretation (\cite{INSF}).

Interpret graphs as cobordisms and extensions as composition of cobordisms
determined by the insertion vertex and the order of matching the external legs.
If $v$ is an internal vertex of $\Gamma_1$ for instance,
the insertion of $\Gamma_2$ at the vertex $v$ (with the additional data $\sigma$
regarding the vertex matching order),
precisely corresponds to the composition of the corresponding cobordisms:
$$\Gamma_1\circ^\sigma_v\Gamma_2=[\Gamma_1-v]\circ[\Gamma_2], \qquad
\emptyset\overset{[\Gamma_2]}{\to}[k],\quad [k]\overset{[\Gamma_1-v]}{\to} [m], $$
where $k$ is the valency of $v$, and $\Gamma_1-v$ is the graph
obtained by cutting the vertex $v$ out 
($\Gamma_1-v$ will have both an ``In'' and an ''Out'' boundary).

In this context, 
the Euler-Poincare property of a Feynman rule generalizes in
the present context (states on graphs, i.e. graph cohomology) to a ``propagator property'':
$$K(In,Out)=\sum_{states\ \phi}K(In, \phi)K(\phi,Out).$$
Since here the propagator is essentially $<g|U_\Gamma(\phi)|f>$
(if $\Gamma$ has both an In and Out boundary),
one may chose to consider extensions at the level of states (graph cohomology):
$$(\Gamma_2,\phi_2)\hookrightarrow(\Gamma,\phi)\twoheadrightarrow(\Gamma_1,\phi_1).$$
%
%
Here $\phi_2$ and $\phi_1$ are determined as restrictions of $\phi$
to $\Gamma_2$ and $\Gamma_1-v$,
while the state of the vertex $v\in\Gamma_1$ is the ``effective state'' of $\Gamma_2$:
$$\phi_1(v)=U_{\Gamma_2}(\phi_2)$$
(see operation $\bullet$ and Lemma \ref{L:bullet} below).

The following basic property of $U$ is expected
(generalized Euler-Poincare map / propagator property).
\begin{prop}\label{P:pp}
If $(\Gamma_2,\phi_2)\hookrightarrow(\Gamma,\phi)\twoheadrightarrow(\Gamma_1,\phi_1)$, then:
$$U_\Gamma(\phi)=U_{\Gamma_1}(\phi_1)\circ U_{\Gamma_2}(\phi_2).$$
\end{prop}

%
%
\subsection{The generalized Feynman rule}
Returning to our main objective,
we still have to prove that:
\begin{prop}\label{P:prelie}
$U$ is a pre-Lie morphism:
$$U_{\Gamma_1\star_b\Gamma_2}=U_{\Gamma_1}\circ U_{\Gamma_2},$$
where the extensions defining the product $\star_b$ correspond to subgraphs
intersecting the boundary.
\end{prop}
\begin{pf}
The above claim is a consequence of the more basic fact regarding 
insertions of appropriate graphs at a vertex.
The ``Gerstenhaber-like'' compositions from the above right hand sides
are typical for this TQFT gluing/composition operations,
as sketched in Section \ref{S:TQFT}.
One would then establish the claim at the level of the corresponding TQFT,
using the propagator property (Proposition \ref{P:pp}).
\end{pf}
It is not clear for the moment the role of the above restriction to boundary meeting
extensions (discarding the ``vacuum fluctuations'').

Note also the following relation with the operation $\bullet$ on polyvector fields.
\begin{lem}\label{L:bullet}
If $\Gamma'=\Gamma/e$ is obtained by collapsing an edge of $\Gamma$, then:
$$U_{\Gamma'}\circ U_e(\gamma)=\sum_{i\ne j}U_{\Gamma'}
((\gamma_i\bullet\gamma_j)\wedge...).$$
\end{lem}
\begin{pf}
Again this is a consequence of the above ``propagator property'' (Proposition \ref{P:pp})
where the state on the collapsed edge is:
$$\phi(v)=\gamma_i\bullet\gamma_j.$$
\end{pf}
The following consequence is claimed (see also \cite{Kon1}, 6.4.1.1., p.25).
\begin{cor}\label{C:bullet}
For all $\Gamma'\in\C{G}$:
$$U_{\Gamma'\star e}(\gamma)=\sum_{i\ne j}U_{\Gamma'}((\gamma_i\bullet\gamma_j)\wedge...).$$
\end{cor}
Perhaps one can avoid involving the pre-Lie operation,
and remain at the level of Lie algebras/UEAs ($L_\infty$-algebras).

%
\begin{rem}
If $B_{\Gamma}=U_\Gamma(\alpha\wedge...\wedge\alpha)$ 
where $\alpha$ is the Poisson structure \cite{Kon1}, p.28,
then $B_\Gamma$ is a Feynman integral corresponding to the propagator $\alpha$.
\end{rem}

%
\section{Applications}\label{S:appl}
As a first application of the previous formalism,
we interpret Kontsevich formality between 
the two DGLAs $T=T_{poly}(\D{R}^d)$ and $D=D_{poly}(\D{R}^d)$ of \cite{Kon1}.
We will prove that the $L_\infty$-condition (F) from \cite{Kon1}, p.24:
\begin{alignat}{2}\label{E:lim2}
&(F1) \quad &\sum_{i\ne j}\pm 
    U_{n-1}((\gamma_i\bullet\gamma_j)\wedge...\wedge\gamma_n)\\
&(F2) \quad +&\sum_{k+l=n}1/(k!l!)\sum_{\sigma\in\Sigma_n}\pm U_k\circ U_l(\gamma_\sigma)=0,
\end{alignat}
follows in a direct way from the fact that $U$ is a generalized Feynman integral,
and it preserves the pre-Lie composition of Feynman graphs (Definition \ref{D:extprod}),
as claimed in the previous section.
This will essentially yield a proof of 
the general result of \S\ref{S:linf} (Theorem \ref{T:f1}).
\begin{th}\label{T:dgla}
(i) $[Q,U]=(\delta W) U$, 
where $Q$ denotes the appropriate $L_\infty$-structure;

(ii) $U$ is an $L_\infty$-algebra morphism iff $\delta W=0$.
\end{th}
\begin{pf}
We will prove (i), 
since (ii) becomes clear after recalling that $U$ is an $L_\infty$-morphism
iff $[Q,U]=0$ (\cite{Kel}) or $(F)$ holds true \cite{Kon1}.
Here $[Q,U]=Q_1\circ U\pm U\circ Q_2$ (see \cite{Kel}, 8,9,12),
and $\delta W(\Gamma)=W(\Delta\Gamma)=W(d^{int}\Gamma)+W(\Delta_b\Gamma)$.
Instead of $[Q,U]$ we will refer to its alternative form $(F)$.

As stated in \cite{Kon1},
the $L_\infty$-algebra condition (F) 
corresponds to $W(d^{int}\Gamma)$ (first line - 6.4.1.1, p.25) 
and $W(\Delta_b\Gamma)$ (second line - 6.4.2.1., p.26).
Since the other integrals vanish,
the sum of the two contributions equals $W(\Delta)$.

Indeed, we will compute the coefficients $c_\Gamma$, 
and prove that:
\begin{equation}\label{E:coeffs}
c_\Gamma=\delta W(\Gamma).
\end{equation}
Recall that $U=\sum_nU_n$ and 
$U_n=\sum_{m\ge 0}\sum_{\Gamma\in G_{n,m}^{-1}}W_\Gamma U_\Gamma,$
where $G_{n,m}^{-1}$ is the set of graphs with $n$ internal vertices, 
$m$ external vertices and $2n+m-2$ edges.
Therefore $U_\Gamma:T^n\to D_m$, 
and $U_\Gamma(\gamma):A^m\to A$, where $A=C^\infty(M)$.

Substitute the above Feynman expansion in equation (F), to obtain:
\begin{align}
&\sum_{\Gamma'\in G_{n-1,m}}\pm W_{\Gamma'}\sum_{i\ne j}U_{\Gamma_1}
((\gamma_i\bullet\gamma_j)\wedge...) \tag{F1}\\
+&\sum_{\Gamma_1\in G_{k,m}, \Gamma_2\in G_{l,m}}\pm W_{\Gamma_1}W_{\Gamma_2}
U_{\Gamma_1}\overset{\circ}{\wedge}U_{\Gamma_2}=0,\tag{F2}
\end{align}
where the result of alternating the Gerstenhaber composition was denoted by:
$$U_{\Gamma_1}\overset{\circ}{\wedge}U_{\Gamma_2}(\gamma)=
1/(k!l!)\sum_{\sigma\in\Sigma_n}
U_{\Gamma_1}(\gamma_{\sigma(1)}\wedge...)\circ
U_{\Gamma_2}(\gamma_{\sigma(k+1)}\wedge...).
$$
In order to compare it with our claim (Equation \ref{E:coeffs}):
\begin{equation}\label{E:coef}
c_\Gamma=\sum_{\Gamma_1\to\Gamma\to\Gamma_2}\pm W_{\Gamma_1}W_{\Gamma_2},
\end{equation}
use the above lemmas and rearrange the sums.
The first line (F1) transforms as follows:
\begin{align}
\sum_{\Gamma'\in G_{n-1,m}}W_{\Gamma'}\sum_{i\ne j}U_{\Gamma'}((\gamma_i\bullet\gamma_j)\wedge...)
&=\sum_{\Gamma'}W_{\Gamma'} U_{\Gamma'\star e}(\gamma), 
  \quad by\ Corollary\ \ref{C:bullet}\quad \\
&=\sum_{e\hookrightarrow\Gamma\twoheadrightarrow\Gamma',\ e\in\Gamma^{(1)}_{int}}
  W_{\Gamma'}U_\Gamma(\gamma)\quad by\ Definition\ \ref{D:extprod}\\
&=\sum_{\Gamma\in G_{n,m}}(\sum_{e\hookrightarrow\Gamma\twoheadrightarrow\Gamma',\ e\in\Gamma^{(1)}_{int}}
  \pm W_{\Gamma/e})
U_\Gamma(\gamma)\\
&=\sum_{\Gamma\in G_{n,m}}W_{d^{int}\Gamma}U_\Gamma(\gamma)\quad by\ Equation\ \ref{E:fgdiff}.
\end{align}
To transform the second line (F2), 
postpone the application of the alternation operator $\wedge$:
\begin{align}
\sum_{\Gamma_1\in G_{k,m}, \Gamma_2\in G_{l,m}}
& \pm W_{\Gamma_1}W_{\Gamma_2} U_{\Gamma_1}\circ U_{\Gamma_2}
= \sum_{\Gamma_1,\Gamma_2}\pm W_{\Gamma_1}W_{\Gamma_2}
  U_{\Gamma_1\star_b\Gamma_2}\quad by \ Proposition\ \ref{P:prelie}\\
&= \sum_{\Gamma_1,\Gamma_2}W_{\Gamma_1}W_{\Gamma_2}\
   \sum_{\Gamma_1\hookrightarrow\Gamma\twoheadrightarrow\Gamma_2,
   \ \Gamma_1\cap\partial\Gamma\ne\emptyset}
  \pm U_{\Gamma}\quad \sim \ Definition\ \ref{D:extprod}\\
&= \sum_{\Gamma_1,\Gamma_2}\quad
   \sum_{\Gamma_1\hookrightarrow\Gamma\twoheadrightarrow\Gamma_2, 
     \ \Gamma_1\cap\partial\Gamma\ne\emptyset}
  \pm W_{\Gamma_1}W_{\Gamma_2} U_{\Gamma}\\
&=\sum_{\Gamma\in G_{n,m}} W(\Delta_b\Gamma) U_\Gamma,
\end{align}
where Propositions \ref{P:multiplic} and Proposition \ref{P:s2} were used.
Adding the two expressions yields the nonzero terms from 
the right hand side of Equation \ref{E:coef}.
\end{pf}
\begin{rem}
Working with the above equalities after applying the alternation operator amounts 
to proving the statements at the level of Lie algebras,
avoiding the pre-Lie operations.
\end{rem}
\begin{rem}
The equation (i) from the Theorem \ref{T:dgla} may be interpreted as:
$$``ad_Q(U)''=(\delta W)U,$$
i.e. that pre-$L_\infty$-morphisms which are Feynman expansions
are solutions of an ``eigenvalue problem'':
$U$ is an eigenvector corresponding to the eigenvalue $\delta W$.
The ``kernel'' consists of $L_\infty$-morphisms.
\end{rem}
%
The formality is obtained as a corollary.
\begin{cor}\label{C:DGLAf}
The pre-$L_\infty$-morphism $U:T\to D$ is an $L_\infty$-morphism.
\end{cor}
\begin{pf}
By Corollary \ref{C:Fcocycle} the Feynman integral $W$ determined by the
configuration functor $(C,\bar{C},\partial)$ is a cocycle.
To conclude apply the above theorem.
\end{pf}
As a second application we envision an algebraic/combinatorial Feynman integral 
(closer in spirit to the ``book keeping'' of a Gaussian expansion)
based on the Hopf algebra of forests
(``labels of the boundary of $S_\Gamma$'' - see \cite{Kon1}, p.20).
More generally, 
the problem (to be investigate elsewhere) is to find suitable 
examples of pairings with Lagrangians, 
yielding such cocycles.
Then the corresponding system of weights would provide
a formula for the star-product of a Poisson manifold \cite{Kon1}.

\section{Conclusions and further developments}\label{S:c}
The intent of the present article is to isolate some algebraic properties of,
and to establish a perhaps simpler ``interface'' to a mathematical model for 
the Feynman path integral quantization based on homotopical algebra:
$$"\quad \int \C{D}\gamma\ e^{S[\gamma]}\quad "=>\sum_n\sum_\gamma U_n(\gamma).$$
The left hand side is a conceptual framework which need not be implemented using  
analytical tools (integrals, measures, etc.), 
but most likely with algebraic tools, e.g. state sum models yielding TQFTs etc..
The perturbative approach guided by a formal LHS implements an expansion which is still
formal, requiring renormalization.

BV-formalism and $L_\infty$-formalism are two such approaches
to perturbative QFT (\cite{BV,AKSZ}).
The geometric approach of the BV-formalism implements the framework
of Feynman path integral quantization method, 
starting with a classical action, 
finding the quantum BV-action 
$S_{BV}=S_{free}+S_{int}$, 
fixing the gauge and applying a reduction technique (e.g. Faddeev-Popov)  \cite{Cat1,Cat2}, 
and finally expanding the action in perturbation series
labeled by Feynman diagrams (see \cite{Riv} for a concise introduction).

The $L_\infty$-algebra approach is a direct implementation of the RHS.
>From this point of view,
to write a Lagrangian in the context of $L\infty$-algebras \cite{Wit}, p.695,
could reasonably mean that a certain formal Lagrangian in the LHS 
is implemented in the $L_\infty$-algebra context of the RHS 
by providing a Feynman category corresponding to the interaction Lagrangian, 
and a Feynman rule correspond to the free field theory.
The Green functions would be obtained as Feynman coefficients of 
$L_\infty$-morphims expanded over a given class of Feynman graphs.
In other words, 
the partition function understood as a generating function
for the Green functions, 
may be implemented as an $L_\infty$-morphism.

At a principial level one can defend the above strategy, by claiming that the RHS is
conceptually closer to the spirit of quantum theory focusing on describing correlations.
The philosophy sketched in \cite{Irem} 
reinterprets the concept of space-time as a 
receptacle of interactions/transitions between states,
and adequately modeled by ``categories with Lagrangians'',
while the LHS comes from the traditional ``manifold approach to space-time''
trying to force integrals in the sense of analysis ``converge''.

The former philosophy can be implemented by defining a ``Feynman category''
to be essentially a ``generalized cobordism category'' (\cite{INSF}),
with actions as functors (see \cite{Irem}).
Cobordism categories and TQFTs, tangles, operads, PROPs, and various other graphical calculi
can be restated in terms of Generalized Cobordism Categories and their representations.

\vspace{.1in}
Of course, to provide meaningful Feynman categories, one has to look at the
conceptual framework of FPI (physics side), which, we would like to emphasize it again,
needs not be ``mathematically rigorous'':
\vspace{.1in}
\begin{alignat}{2}
&\underline{Quantum\ Field\ Theory}\qquad       &     &\underline{Feynman\ Path\ Integral\ Quantization}\notag \\
&Interface &     &Geometric \qquad \qquad \qquad Algebraic / Categorical\notag\\
&Implementation &   &BV\quad \to AKSZ \to \quad A_\infty/L_\infty-algebras / categories\notag
\end{alignat}

\vspace{.5in}

{\bf Glossary}

The following table is a (tentative) dictionary of selected terms and notations used.
The symbol $\sim$ is used to denote merely a relation.

\vspace{.1in}
\noindent
\begin{tabular}{|l|l|l|}
\hline
Notation & Mathematics & Physics \\ \hline
$\C{G}$ & Gen. Cobord. Cat. & Class of Feynman graphs \\ \hline
$\Gamma$ & Object & Feynman graph \\ \hline
$g$ & Lie algebra of PIs & Feynman graphs + insertions \\ \hline
$H(\C{G})$ & Grothendieck DG-coalgebra & Feynman graphs\\ \hline
$\C{F}_\bullet$ & $L_\infty$-morphism & Partition function \\ \hline
$\C{F}_n$ & n-th derivative & Green function \\ \hline
$W$ & Weight / cocycle & $\sim$ Feynman integral \\ \hline
$U$ & pre-Lie algebra morphism & $\sim$ Feynman integral \\ \hline
$H(H;k)$ & $L_\infty$-morphisms & Feynman expansions \\ \hline
$S$ & Configuration functor & Configuration spaces \\ \hline
$\omega(\Gamma)$ & Closed top form on $S(\Gamma)$ & Interaction Lagrangian \\ \hline
$W=<S,\omega>$ & Pairing & $\sim$ Action \\ \hline
\end{tabular}



\end{document}